%% file: paper.tex
\documentclass[sigplan,10pt, nonacm]{acmart}
\input{packages}

\input{macros}

\begin{document}

% paper information
\title[Porcupine]{Porcupine: A Synthesizing Compiler \\for Vectorized Homomorphic Encryption}

\author{Meghan Cowan}
\affiliation{
  \department{Facebook Reality Labs Research}
}
\email{meghancowan@fb.com}

\author{Deeksha Dangwal}
\affiliation{
  \department{Facebook Reality Labs Research}
}
\email{ddangwal@fb.com}

\author{Armin Alaghi}
\affiliation{
  \department{Facebook Reality Labs Research}
}
\email{alaghi@fb.com}

\author{Caroline Trippel}
\affiliation{
  \department{Stanford University}
}
\email{trippel@stanford.edu}

\author{Vincent T. Lee}
\affiliation{
  \department{Facebook Reality Labs Research}
}
\email{vtlee@fb.com}

\author{Brandon Reagen}
\affiliation{
  \department{New York University}
}
\email{bjr5@nyu.edu}

% Short author headers
\renewcommand{\shortauthors}{M.Cowan, D.Dangwal, A.Alaghi, C.Trippel, V.T. Lee, and B. Reagen}

% document text
\input{text/00-abstract}

\maketitle

\input{text/01-introduction}
\input{text/02-background}
\input{text/03-problem}
\input{text/04-synthesis}

\input{text/05-engine}

\input{text/06-optimizations}
\input{text/07-evaluation}
\input{text/08-related-work}
\input{text/09-conclusion}

% citations
\balance
\bibliographystyle{ACM-Reference-Format}
\bibliography{references}

\end{document}

%% file: packages.tex
\usepackage{algorithm}
\usepackage{algpseudocode}
\usepackage{balance}
\usepackage{booktabs}
\usepackage{bookmark}
\usepackage{comment}
\usepackage{datetime}
\usepackage{enumerate}
\usepackage{fancyhdr}
\usepackage{framed}
\usepackage{graphicx}
\usepackage{hyperref}
\usepackage[procnames]{listings}
\usepackage{microtype}
\usepackage{multirow}
\usepackage{subcaption}
\usepackage{textgreek}
\usepackage{enumitem}
\usepackage{url}
\usepackage{xcolor}
\usepackage{xspace}

%% file: macros.tex
\newcommand{\armadillo}{Porcupine\xspace}
\newcommand{\sketchname}{local rotate\xspace}
\newcommand{\code}[1]{\texttt{#1}}

\graphicspath{{figures/}}
\DeclareGraphicsExtensions{.pdf,jpeg,.png}

\newcommand{\speedupbest}{51\%\xspace} % Gy
\newcommand{\speeduppoly}{27\%\xspace}
\newcommand{\speedupsobel}{6\%\xspace}
\newcommand{\speedupharris}{13\%\xspace}

\lstdefinestyle{mystyle}{
    basicstyle=\ttfamily\small,
    breakatwhitespace=false,         
    breaklines=true,                 
    captionpos=b,                    
    keepspaces=false,                 
    numbersep=2pt,                  
    showspaces=false,                
    showstringspaces=false,
    showtabs=false,                  
    tabsize=2
}
\lstdefinelanguage{rosette}{
  morekeywords=[1]{verify,solve,forall,and,or,assert,s-exp,rosette,set!,begin,define,define-values,define-syntax,define-syntax-rule,syntax-rules,let,let*,if,when,unless,match-define,lambda,provide,cond,case,else,struct,letrec,for/list,true,false,null,local,require,rename-in,??,define-symbolic,define-symbolic*,not,=>,ite,\#lang,\#:transparent,\#:mutable,equal?,match,list,for},
  morekeywords=[2]{bv,bvult,bvuge,bvsub,~>,bitvector,integer?},
  morekeywords=[3]{serval:split-pc,serval:bug-on},
  alsoletter={\#,:,?,-,=>},
  morecomment=[l]{;},
}

%% file: text/00-abstract.tex
\begin{abstract}

Homomorphic encryption (HE) is a privacy-preserving technique
that enables computation directly on encrypted data.
Despite its promise,
HE has seen limited use due to performance overheads and
compilation challenges.
Recent work has made significant advances to address the performance
overheads but automatic compilation of efficient HE kernels remains relatively unexplored.

This paper presents \textit{\armadillo} --- 
an optimizing compiler --- and HE DSL named \textit{Quill} to
automatically generate HE code using program synthesis.
HE poses three major compilation challenges:
it only supports a limited set of SIMD-like operators,
it uses long-vector operands,
and decryption can fail if ciphertext noise growth is not managed properly.
Quill captures the underlying HE operator behavior
that enables Porcupine to reason about the complex
trade-offs imposed by the challenges and generate
optimized, verified HE kernels.
To improve synthesis time, we propose a series of optimizations 
including a sketch design tailored to HE and instruction restriction to
narrow the program search space.
We evaluate \armadillo using a set of kernels
and show speedups of up to \speedupbest (11\% geometric mean)
compared to heuristic-driven hand-optimized kernels.
Analysis of \armadillo's synthesized code reveals
that optimal solutions are not always intuitive,
underscoring the utility of automated reasoning in this domain.

\end{abstract}

%In HE, each ciphertext has an initial noise level that increases with successive operations;
%if the noise grows too much, decryption fails and the result is unusable.
%This challenge is compounded by the SIMD instruction set which requires careful packing
%and kernel generation to pack kernels across vector slots and correctly build a HE kernel.
%The SIMD nature of HE instructions also poses a
%challenge for compilers that must expertly schedule instructions,
%align packed operands/vector slots,
%and satisfy HE noise budgets to generating performant and correct programs.

%\meghan{we would need to introduce rotation here? or keep rotation vague as ciphertext data movement?} - added something here
%A user provides a program specification and sketch and \armadillo automatically generates packed vectorized HE kernels using program synthesis.
% \armadillo further improves \fixme{performance} and synthesis time
% via a series of \fixme{novel} optimizations. % to \fixme{x, y, z}.

%% file: text/01-introduction.tex
\section{Introduction}
\label{sec:introduction}

Homomorphic encryption (HE) is a rapidly maturing privacy-preserving technology
that enables computation directly on encrypted data.
HE enables secure remote computation, as cloud service providers can compute on data without viewing private actual content.
Despite its appeal, two key challenges prevent widespread HE adoption:
performance and programmability.
Today, most systems-oriented HE research has focused on overcoming the prohibitive
performance overheads with high-performance
software libraries~\cite{palisade, sealcrypto} and custom hardware~\cite{cheetah, heax}.
The performance results are encouraging with some suggesting that 
HE can approach real-time latency for certain applications with sufficiently large hardware resources~\cite{cheetah}.
Realizing the full potential of HE requires an analogous compiler effort to alleviate
the code generation and programming challenges, which remain less explored.

Modern ring-based HE schemes pose three programming challenges:
(i) they only provide a limited set of instructions (add, multiply, and rotate);
(ii) operands are long vectors, on the order of thousands;
(iii) ciphertexts have noise that grows as operations are performed
and causes decryption to fail if too much accumulates.
For instance,  Brakerski/Fan-Vercauteren (BFV) crpytosystem~\cite{bfv} operates on vectors that
\textit{packing} multiple data into a single ciphertext to improve performance.
Instructions operating on packed-vector ciphertexts can be abstracted as a SIMD
(single instruction, multiple data) instruction set,
which introduces vectorization challenges.

To target the instruction set, the programmer must break down an input kernel into SIMD addition, multiply, and rotation instructions,
while minimizing noise accumulation.
These challenges introduce a complex design space when implementing HE kernels.
As a result, HE kernels are currently written by a limited set of experts fluent
in ``HE-assembly'' and the details of ciphertext noise.
Even for experts, this process is laborious.
As a result, hand-writing HE programs does not scale beyond a few kernels.
\textit{Thus, automated compiler support for HE is needed for it to emerge
as a viable solution for privacy-preserving computation.}

A nascent body of prior work exists and has investigated
specific aspects of compiling HE code.
For example, prior work has shown HE parameter tuning, which determines the noise budget,
can be automated and optimized to improve performance~\cite{chet, eva, ramparts, cingulata}.
Others have proposed mechanisms to optimize data layouts for neural networks~\cite{chet}.
Prior solutions have also used a mix of symbolic execution~\cite{ramparts} and rewrite rules~\cite{eva, cingulata, ngraph} for 
code generation and 
optimizations for logic minimization 
(e.g., Boolean logic minimization~\cite{cingulata,lee2020optimizing}. 
Each of these lines of work have advanced the field and addressed notable HE compilation challenges.
In contrast to related work (see~\autoref{sec:related_work}),
we are the first to automate compiling and optimizing vectorized HE kernels.

In this paper we propose Porcupine, a synthesizing compiler for HE.
Users provide a reference of their plaintext kernel and \armadillo synthesizes a vectorized HE kernel that performs the same computation. 
Internally, \armadillo models instruction noise, latency, behavior, and HE program semantics with 
Quill: a novel HE DSL.
Quill enables Porcupine to reason about and search for HE kernels that are 
(verifiably) correct and minimizes the kernel's cost, i.e., latency and noise accumulation.
With \armadillo and Quill, we develop a synthesis procedure that automates 
and optimizes the mapping and scheduling of plaintext kernels to HE instructions.

\armadillo uses syntax-guided synthesis~\cite{syntax-guided-synthesis} so that our synthesizer completes a sketch, or HE kernel template.
We introduce a novel \textit{\sketchname} that treats ciphertext rotation as an input to HE add and multiply instructions rather than an independent rotation instruction; this makes the synthesis search more tractable by limiting the space of possible programs. 
Furthermore, we develop several HE-specific optimizations including rotation restrictions for tree reductions and stencil computations, multi-step synthesis, and constraint optimizations
to further improve synthesis run time (details in~\autoref{sec:scaling_up}). 

We evaluate \armadillo using a variety of image processing and linear algebra kernels.
Baseline programs are hand-written and attempt to minimize logic depth, the current best practice for optimizing HE programs~\cite{ramparts, cingulata, lee2020optimizing}.
For small kernels, \armadillo is able to find the same optimized implementations as the hand-written baseline.
On larger, more complex kernels, we show \armadillo's programs are up to \speedupbest faster.
Upon further analysis, we find that \armadillo can discover optimizations such as
factorization and even application-specific optimizations involving separable filters.
Our results demonstrate the efficacy and generality of our synthesis-based compilation approach
and further motivates the benefits of automated reasoning in HE for both performance and productivity.

This paper makes the following contributions:
\begin{enumerate}

\item We present \armadillo, a program synthesis-based compiler that
      automatically generates vectorized HE programs, and
      Quill, a DSL for HE.
      \armadillo includes a set of optimizations needed to effectively
      adopt program synthesis to target HE.

\item We evaluate \armadillo using nine kernels to demonstrate it can
      successfully translate plaintext specifications to correct HE-equivalent implementations.
      \armadillo achieves speedups of up to \speedupbest (11\% geometric mean)
      over hand-written baselines implemented with best-known practices.
      We note situations where optimal solutions cannot be found with existing techniques
      (i.e., logic depth minimization),
      further motivating automated reasoning-based solutions.

\item We develop a set of optimizations to improve Porcupine’s 
        synthesis time and compile larger programs.
        First, we develop a domain-specific \sketchname that
         considers rotations as inputs to arithmetic instructions, narrowing the
         solutions space without compromising quality.
         We further restrict HE rotation patterns and
         propose a multi-step synthesis process.

\end{enumerate}

%% file: text/02-background.tex
\section{Homomorphic Encryption Background}
\label{sec:background}
This section provides a brief background on homomorphic encryption. 
We refer the reader to~\cite{gentry10, bfv, bgv, bos2013improved, brakerski2012fully} for the more technical details of how HE works.

\begin{figure}[t]
\centering
\includegraphics[width=\linewidth]{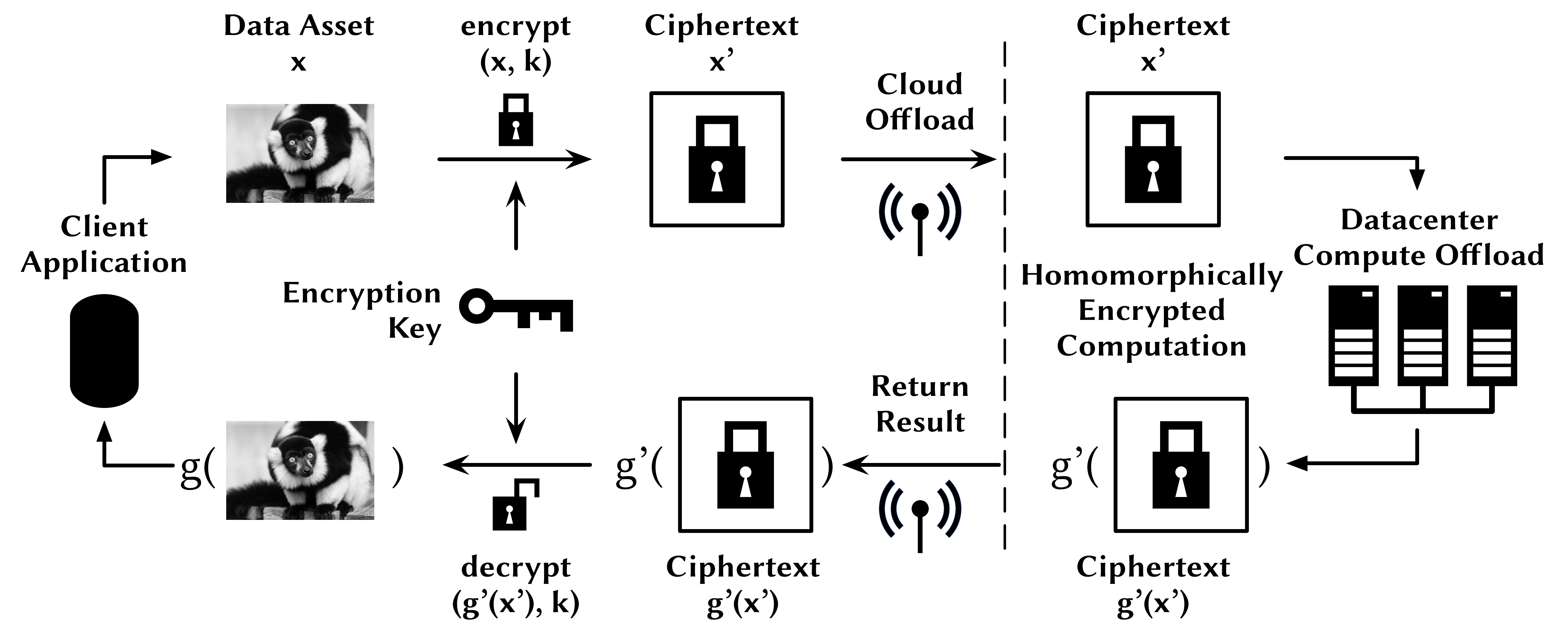}
\caption{HE system for to an
  untrusted third party cloud. A plaintext data asset $x$ is
  encrypted with a key $k$ to generate ciphertext $x'$ and transmitted to the cloud. The cloud service
  applies HE computation $g'$ to the ciphertext \emph{without} decrypting the
  data. The result $g'(x')$ is transmitted back to client where decryption yields the result $g(x)$.
}
  \label{fig:he_system_diagram}
\end{figure}

\subsection{Homomorphic Encryption Basics}

Homomorphic encryption enables arbitrary computation over encrypted data or ciphertexts ~\cite{gentry09}. 
This enables secure computation offload where an untrusted third party, such as a cloud provider, performs computation over a client's private data without gaining access to it.

\autoref{fig:he_system_diagram} shows a canonical HE system for secure cloud compute.
First, the client locally encrypts their data asset $x$ using a private key $k$.
The resulting ciphertext $x'$ is then sent to the cloud where an HE function $g'$ is applied to it.
The output of the computation $g'(x')$ is then sent back to the client
and decrypted using the same key $k$ to reveal the plaintext output: $g(x)$.
HE allows us to define a function $g'$ that operates over ciphertext $x'=\textrm{encrypt}(x, k)$ such that:
\[\textrm{decrypt}(g'(x'), k) = g(x)\]

The private key $k$ never leaves the client,
ensuring the client's data asset is secure throughout the computation.
Additionally, the client does not learn $g$,
which could be a secret that the cloud wants to protect.
\armadillo's goal is to synthesize $g'$ given a definition of the kernel $g$. 

This paper focuses on the BFV cryptosystem, a specific HE scheme that targets integers~\cite{bfv}.
In the remainder of this section, we provide an overview of the BFV scheme and focus on the vector programming model, instructions, and noise considerations it exposes.
For a more technical description see ~\cite{bfv,he-standard}.

\subsection{BFV}
BFV is an HE scheme that operates over encrypted integers.
In BFV, integers are encrypted into a ciphertext polynomial of degree $N$ with integer coefficients that are modulo $q$.
A key property of BFV is batching;
this allows a vector of up to $N$ integers to be encrypted in a single ciphertext with operations behaving in a SIMD manner.

For the most part, ciphertext polynomials behave as a vector of $N$ slots with bitwidth $q$.
$N$ and $q$ are BFV HE parameters set to provide a desired security level and computational depth,
not the number of raw integers that are encrypted. 
Regardless of whether we encrypt a single integer or $N$ integers in a ciphertext, a vector of $N$ slots is allocated for security purposes. 
$N$ is required to be a large power of two and is often in the tens of thousands,
which makes batching crucial to efficiently utilizing ciphertext space.

\paragraph{Instructions.}
BFV provides three core ciphertext instructions that behave like element-wise SIMD instructions: SIMD add, SIMD multiply, and SIMD (slot) rotate.
Additionally, BFV supports variants of add and multiply that operate on a ciphertext and plaintext instead of two ciphertexts. 

Consider two vectors of integers
$X = \{x_0, x_1, ..., x_{n-1}\}$ and $Y = \{y_0,y_1, ..., y_{n-1}\}$
with ciphertext representation $X'$ and $Y'$ respectively.
SIMD add and multiply both perform element-wise operations over slots.
SIMD add computes $add(X', Y')$ such that
$\textrm{decrypt}(add(X', Y)'), k) = \{x_0 + y_0, x_1 + y_1, ..., x_{n-1} + y_{n-1}\}$, where $k$ is the key used for encryption of $X'$ and $Y'$.
Similarly, the SIMD multiply instruction processes $mult(X, Y)$
so that $\textrm{decrypt}(g'(X', Y'), k) = \{x_0 \times y_0, x_1 \times y_1, ..., x_{n-1} \times y_{n-1}\}$.
Note that the underlying operations that implement $add(X', Y')$ and $mult(X', Y')$
over the ciphertext representations are
\emph{not} simple vector addition or multiplication instructions.

\paragraph{Rotate.}
Additionally, HE provides rotate instructions that circularly shift slots in a ciphertext by an integer amount (similar to bitwise rotations).
Rotations occur in unison: given a rotation amount, all slots shift by the same amount in the same direction and the relative ordering of slots is preserved.
For example, rotating a ciphertext 
$X' = \{x_0,$ $x_1,$ $x_2,...,$ $x_{n-1}\}$ by one element to the left returns 
$\{x_1,$ $x_2,...,$ $x_{n-1},$ $x_0\}$.

Note the ciphertext is not a true vector,
so slots cannot be directly indexed or re-ordered.
Slot rotation is necessary to align slot values between vectors because add and multiply instructions are element-wise along the same slot lanes.
For example, reductions that sum many elements within a ciphertext will need to rotate slots so that elements can be summed in one slot.
Arbitrary shuffles also have to be implemented using rotates and multiplication with masks which can require many instructions and quickly become expensive to implement.

\paragraph{Noise.}
During encryption ciphertexts are injected with random noise to prevent threats such as replay attacks~\cite{replay-attacks}.
During computation this noise grows.
The ciphertext bitwidth $q$ needs to be large enough to contain this noise growth or else the ciphertext becomes corrupted and upon decryption returns an random value (i.e., garbage value).
However, larger values of $q$ increase the memory footprint of ciphertext and requires more compute resource to perform the larger bitwidth arithmetic calculations that back HE instructions.

Specifically, add and rotate additively increase noise, and multiplication multiplicatively increases noise.
Because multiplication dominates noise growth, the multiplicative depth of a program can be used as a guide to select $q$ or as a minimization target.

%% file: text/03-problem.tex
\section{HE Compilation Challenges}
\label{sec:overview}

\begin{figure}[t]
\centering
\includegraphics[width=0.95\linewidth]{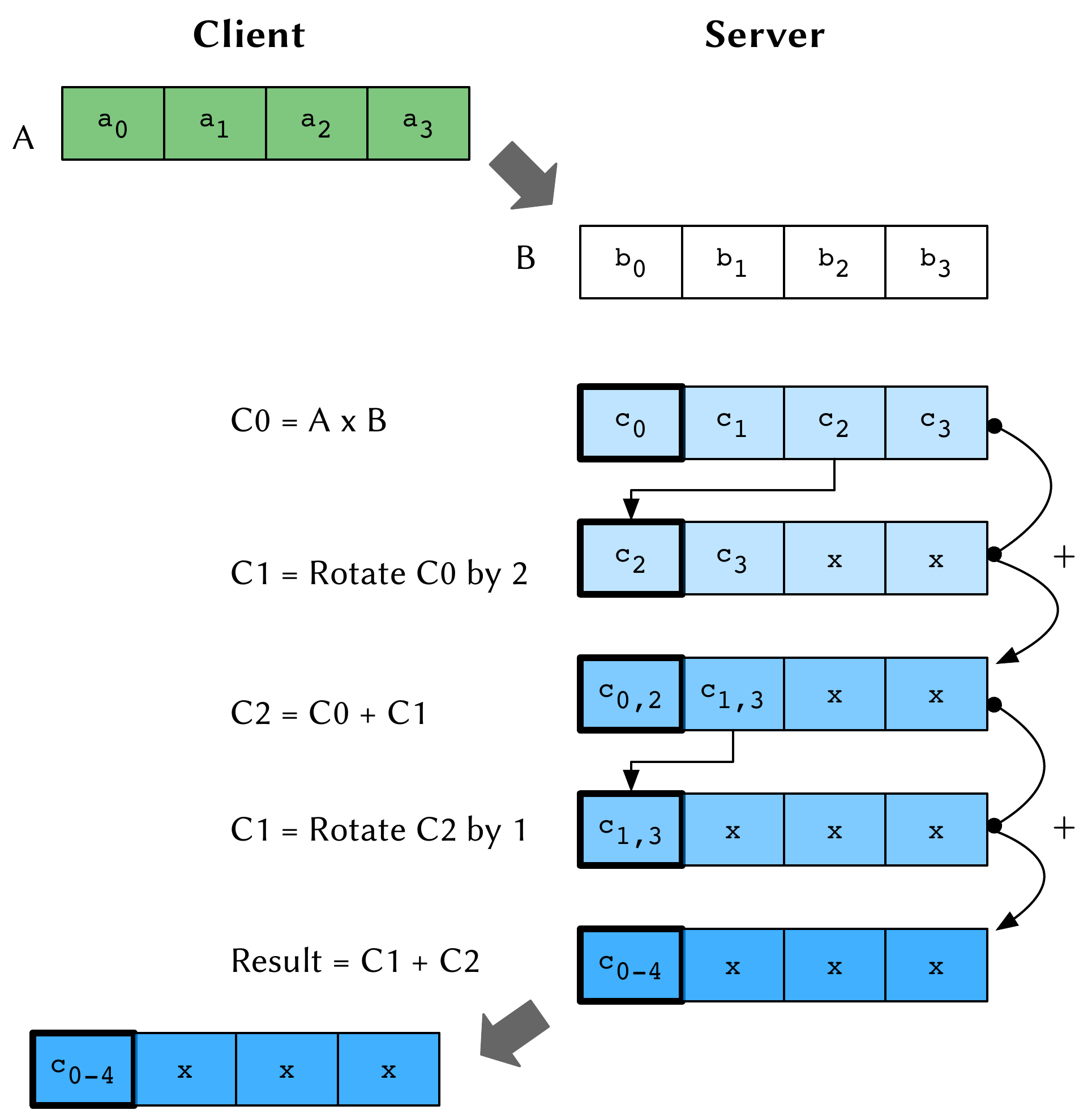}
\caption{HE vectorized dot product implementation.
Given an encrypted input from the client (A),
the server performs an element-wise 
multiplication with server-local data (B).
A reduction is performed using a combination of rotation and
add instructions.
The resulting ciphertext is then
returned the client for decryption.}
  \label{fig:vector_program}
\end{figure}

Handwriting efficient HE kernels is a tedious and error-prone process as
HE provides limited instructions, 
intra-ciphertext data movement must be done using vector rotation, 
and the noise budget adds additional sources of error.
As a result, HE code is today
is typically written by experts~\cite{chet, cheetah, gazelle}.

\armadillo's goal is to automate the generation of vectorized HE kernels
to lower HE's high barrier-to-entry to non-experts as well as time-to-solution for experts.
This section motivates the need for automated reasoning in HE compilers
using a vectorized dot product (see \autoref{fig:vector_program}) as a running example.

\subsection{Data Packing}
To compute an HE dot product, a client sends an encrypted vector of elements to be computed with a server's vector; the encrypted result is then sent back to the client.
A client could encrypt each element of the input vector into individual ciphertexts, 
but this uses only a single slot of each ciphertext vector, wasting the other slots.
Another solution is to batch $N$ independent tasks into a single ciphertext to 
amortize the cost of the ciphertext and HE program.
However, HE vectors can hold tens of thousands of elements and 
most applications cannot rely on batching of this scale.

Instead, a client can pack the input data vector in a single ciphertext,
as shown in \autoref{fig:vector_program}.
In our example of a four element dot product, this requires only one ciphertext, not four.
\textit{\armadillo assumes kernels operate over packed inputs to efficiently utilize memory.}

\subsection{HE Computation}
One of the key challenges for building optimized HE kernels is breaking down scalar computation to efficiently use the limited HE instruction set.
In ciphertext vectors, the relative ordering of packed data elements is fixed; 
thus, while element-wise SIMD addition and multiplication computation is trivial to implement, scalar-output calculations such as reductions
require proper alignment of slots between ciphertext operands.
The only way to align different slot indices between two ciphertexts 
is to explicitly rotate one of them such that the desired elements are aligned to the same slot.

\autoref{fig:vector_program} illustrates how this is done for an HE dot product reduction operation using packed vectors.
The client's and server's ciphertext operands are multiplied together and reduced to a single value.
The multiplication operation is element-wise,
so it can be implemented with a HE SIMD multiply operation.
However, the summation within the vector must be performed by repeatedly rotating and adding ciphertexts together such that the intermediate operands are properly aligned to a slot in the vector (in this case the slot at index 0).
The rotations and arithmetic operations are interleaved to take advantage of the SIMD parallelism and enable reduction to be computed with only two HE add operations for four elements.

For more complex kernels, simultaneously scheduling computations and vector rotations is non-trivial to implement efficiently.
Arbitrary slot swaps or shuffles (e.g., instructions like \code{\_mm\_shuffle\_epi32}) that change the relative ordering of elements in a vector are even more tedious to implement.
While these arbitrary shuffles can be implemented in HE by multiplying with masks and rotation operations,
this is undesirable since it requires dramatically increasing the multiplicative depth and hence noise budget requirements.

\subsection{Performance and Noise}
The vectorization challenges are further complicated by HE's compound-cost
model that must consider both performance and noise.
Performance and noise costs cannot be reasoned about independently;
the performance cost must be aware of the noise growth since the noise budget 
parameter $q$ defines the bitwidth precision of the underlying mathematical HE instruction implementations.
Thus, a larger $q$ increases the latency cost of each HE instruction.
This means any sort of optimization objective for program synthesis will have to consider both noise and performance together.

%% file: text/04-synthesis.tex
\section{\armadillo Compiler and Synthesis Formulation}
\label{sec:synthesis}

This section introduces the \armadillo compiler, Quill DSL,
and the program synthesis formulation used to optimize HE kernels.

{
\begin{figure*}[t]
\centering
\includegraphics[width=\linewidth]{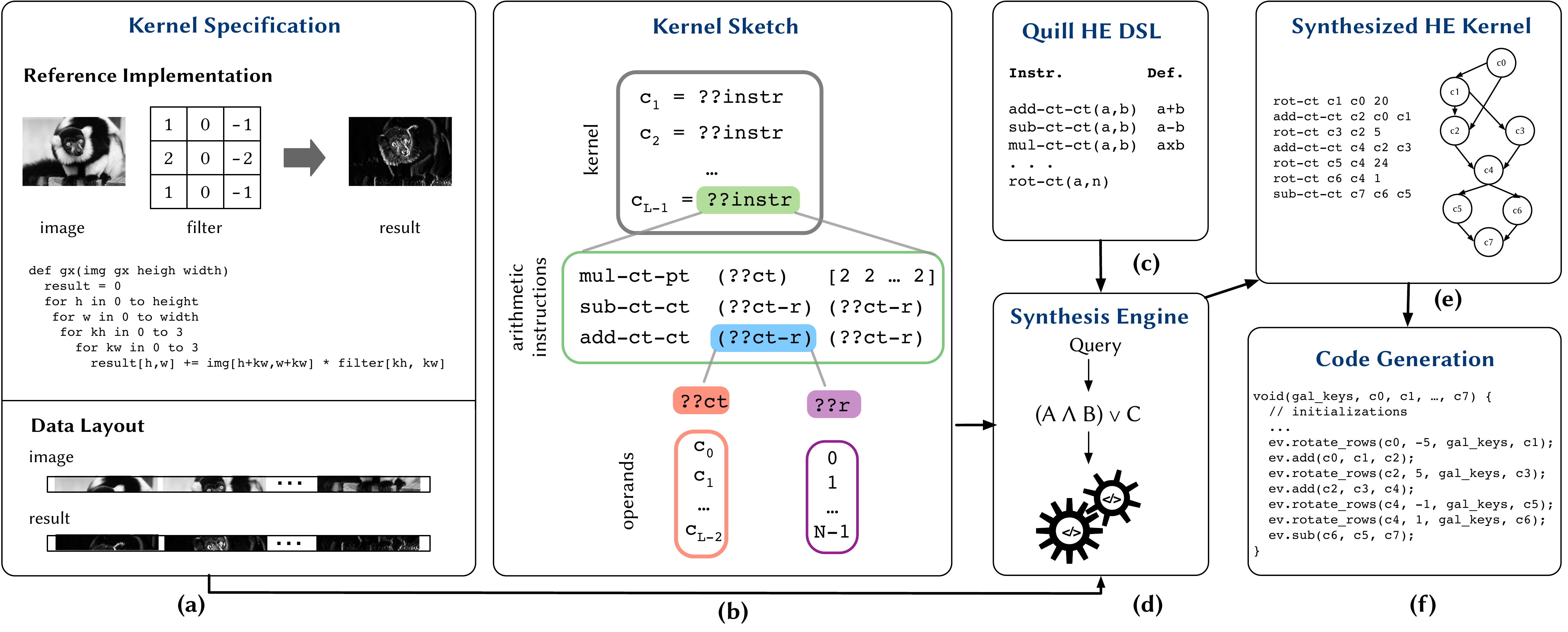}
\caption{The \armadillo compiler. The user provides: (a) a kernel specification and (b) a kernel sketch with ?? denoting holes in the sketch. (c) The Quill DSL encodes the semantics of the HE instruction set and noise models. (d) \armadillo's synthesis engine completes the sketch and synthesizes a program that implements the specification using the Quill DSL. \armadillo uses an SMT solver to automatically solve the vectorization and scheduling challenges so that (e) the synthesized program is optimized. (f) The optimized HE kernel is consumed by code generation to target the  SEAL library~\cite{sealcrypto}.} 
\label{fig:gx-example}
\end{figure*}
}

\subsection{Compiler Overview}

\armadillo is a program synthesis-based compiler that searches for HE kernels rather than relying on traditional rewrite rules.
By searching for programs, \armadillo can discover optimizations that are potentially difficult to identify by hand.
At a high level, \armadillo takes a 
kernel specification (\autoref{fig:gx-example}a) 
and sketch (\autoref{fig:gx-example}b) as input, and 
outputs an optimized HE kernel (\autoref{fig:gx-example}f).
\autoref{sec:he_model} defines our Quill DSL (\autoref{fig:gx-example}c) which is used to model the noise and latency semantics of each HE instruction.
\autoref{sec:specification} defines what composes the specification that \armadillo as input (\autoref{fig:gx-example}a).
\autoref{sec:sketch} explains our sketch formulation and design decisions behind them (\autoref{fig:gx-example}b).
\autoref{sec:synthesis_engine} details our synthesis engine~\cite{syntax-guided-synthesis} which takes the specification, sketch, and HE kernel, and emits a synthesized HE kernel (\autoref{fig:gx-example}\d).

\subsection{Quill: A DSL for HE}
\label{sec:he_model}

% Rewrite
The Quill DSL captures the semantics of HE instructions and their latency-noise behavior,
enabling \armadillo to reason about HE program behavior and noise growth.
Quill is the language that input sketches are written in and 
used by the synthesis engine to infer HE instruction behavior as well as verify correctness.
Quill currently supports BFV~\cite{bfv} HE,
however the techniques are general and can be extended to other ring-based HE schemes,
e.g., BGV~\cite{bgv} and CKKS~\cite{ckks}.

Quill is used to describe straight-line HE programs
that manipulate state initially defined by input vectors (either ciphertext and plaintext).
Quill programs are behavioral models and not true/real HE programs.
The ciphertext operands are implemented as unencrypted vectors
that can only be manipulated according to HE instruction rules, 
which are captured by Quill's semantics.
This provides the benefit that we can compile code without considering the implementation details of true HE.

\paragraph{State in Quill}
In a Quill program, state is defined by plaintext and ciphertext vectors.
All ciphertexts are associated with metadata
that tracks each operand's multiplicative depth, which models noise accumulation.
An input or fresh ciphertext has zero multiplicative depth and 
increases each time a multiplication is performed.
We track only multiplicative depth as it simplifies the objective of noise
minimization without sacrificing accuracy as other instructions - add and rotate -
contribute relatively negligible noise.

\input{text/instructions.tex}

\paragraph{The Quill Instructions.}

Quill supports a SIMD instruction set with a one-to-one mapping to BFV HE instructions.
\autoref{tbl:instructions} describes each instruction's type signature and how they transform state. %and \fixme{state} transformation.
Instructions include addition, multiplication, and rotations of ciphertext instructions as well variants that operate on ciphertext-plaintext operands, 
e.g., multiplication between a ciphertext and plaintext.
Each instruction is associated with a latency derived by profiling
its corresponding HE instruction with the SEAL HE library~\cite{sealcrypto}.

\subsection{Kernel Specification}
\label{sec:specification}

A \textit{specification} completely describes a target kernel's
functional behavior, 
i.e., it defines what the synthesized HE kernel must compute.
In \armadillo, a specification comprises a \textit{reference implementation} of the computation (in plaintext)
and \textit{vector data layout} that inputs and outputs must adhere to.

\paragraph{Reference Implementation.}
Reference implementations are programs written in Racket~\cite{Racket} 
that define the plaintext computation.
We later user Rosette~\cite{rosette} to automatically lift the Racket program to a symbolic input-output expression that defines the program's behavior.
An example reference implementation for the $G_x$ kernel is shown below.
The code takes as input a 2D gray-scale image and calculates the x-gradient by summing and weighting neighboring pixels according to a 3$\times$3 filter.

\begin{lstlisting}[language=rosette]
(define (Gx img height width):
  for h in 0 to height
    for w in 0 to weight:
      for kh in 0 to 3:
        for kw in 0 to 3:
          result[h,w] += img[h+kw, w+kw] * 
                         filter[kh, kw]
    
\end{lstlisting}

\armadillo uses the reference implementation to verify synthesized ones are correct;
the quality of the reference program does not impact synthesized code quality.
As a result, users can focus on writing correct code without
without the burden of performance tuning.

To work correctly, the implementation must describe computation that is directly implementable in HE.
Implementations cannot contain data dependent control flow
such as conditional statements or loops that depend on a ciphertext,
since we cannot see the values of encrypted data.
This is a limitation of HE, and
while it is possible to approximate this behavior, e.g., using a polynomial function,
this is beyond the scope of our work.

\paragraph{Data Layout.}
A data layout defines how the inputs and outputs are packed into ciphertext and plaintext vectors.
In the $G_x$ example, we pack the input and output image into one ciphertext as a 
flattened row-order vector with zero-padding around the borders.
The data layout is an input to the synthesizer only, and
the reference implementation does not need to consider it.
Together, the reference implementation and data layout define the inputs and outputs to the HE program, and \armadillo will synthesize an HE program that achieves that transformation.

\subsection{Sketch}
\label{sec:sketch}

The third input a user provides is a \textit{sketch} which is a template HE kernel written in Quill used to guide 
the synthesis engine towards a solution.
It allows the user to articulate required features of the HE kernel 
to the synthesis engine while leaving other components unspecified as \textit{holes}, 
indicated by \texttt{??}, for the engine to fill in.
The synthesizer then completes the sketch by filling in the holes
to match the functionality of the reference implementation.
We introduce a \textit{\sketchname} sketch to help the user convey hints about ciphertext rotations.
An example of a \sketchname sketch for the $G_x$ kernel is shown below:

\begin{lstlisting}[language=rosette]
; Program sketch of L components
; ct0 is a ciphertext input
(define (Gx-Sketch ct0 L)
    ; choose an existing ciphertext
    (define (??ct)  (choose* ct))
    
    ; choose a rotation amount in range (0,N)
    (define (??r)  
      (apply choose* (range 0 N)))
      
    ; choose a rotation of an existing ciphertext
    (define (??ct-r)
        (rot-ct ??ct ??r))
      
    ; choose an opcode with operand holes
    (for/list i = 1 to L
        (choose*
            (add-ct-ct (??ct-r) (??ct-r))
            (sub-ct-ct (??ct-r) (??ct-r))
            (mul-ct-pt (??ct) [2 2 ... 2]))))
\end{lstlisting}

The sketch describes a kernel template that takes as input a single ciphertext (encrypted image) and applies a kernel composed of \texttt{L} \textit{components} or arithmetic instructions.
In this example the components are: add two ciphertexts, subtract two ciphertexts or multiply a ciphertext by 2.
Each component contains holes for their instruction dependent operands.
Specifically, \texttt{??ct} is ciphertext hole that can be filled with the ciphertext input or a ciphertexts generated by previous components. 
\texttt{??ct-r} is a ciphertext-rotation that introduces two holes: a ciphertext hole and a rotation hole.
The ciphertext hole can be filled with any previously generated ciphertexts and the rotation hole indicates the ciphertext can be rotated by any legal amount ($1$ to $N-1$) or not at all.
Ciphertext-rotation holes indicate the kernel performs a reduction operation over elements and requires rotation to align vector slots.

Writing sketches of this style is relatively simple.
The arithmetic instructions can be extracted from the specification.
In this case add, subtract, and multiplication by 2 were used in the reference implementation.
The set of arithmetic instructions is treated like a multiset of multiplicity $L$, 
and the synthesizer will determine which instructions and how many are needed.
In other words, the sketch does not have to be exact as the synthesizer can choose to ignore instructions;
this once again eases the burden on the user.
Additionally, the user must specify whether instruction operands should be ciphertexts or ciphertext-rotations, and what rotations are allowed.
As a fall back, all ciphertext holes can be made ciphertext-rotation holes; 
however, this will increase solving time as the sketch describes a larger space of programs.

A key feature of our sketches is that we treat rotation as an input to 
arithmetic instructions rather than a component of the sketch.
This is because rotations are only useful when an arithmetic instruction needs to re-align operands;
in isolation, rotations do not perform meaningful computation.
This excludes programs that contain nested rotations since rotations can be combined.
For instance, we disallow (rot (rot c0 1) 2) since this can be more succinctly expressed as (rot c0 3).

% restrictions
The sketches must describe loop-free programs so that Quill can interpret them.
\armadillo requires sketches to be parameterized by the number of components in the program.
\armadillo first explores small (in terms of {\it L}) programs
and iteratively explores larger programs by incrementing {\it L} until a solution is found.

%searching for solutions that use a minimal number of arithmetic instructions,
\paragraph{Solution.}
A \textit{solution} is a completed sketch that matches the reference implementation.
\armadillo's synthesis engine generates solutions by 
filling instruction and operand holes such that the resulting program satisfies the specification
and optimizes the objective functions (minimize instruction count and noise).
The solution \armadillo synthesizes for the above example uses 
three arithmetic instructions and four rotations \footnote{Rotation amounts are adjusted to be relative in example.}:

\begin{lstlisting}[language=rosette]
    c1 = (add-ct-ct(rot-ct c0 -5) c0)
    c2 = (add-ct-ct(rot-ct c1 5) c1)
    c3 = (sub-ct-ct (rot-ct c2 1) (rot-ct c2 -1))
\end{lstlisting}

%% file: text/instructions.tex
{
  \begin{table*}[t]
  \centering
    \caption{Quill instructions and their affect on the data (denoted by $.data$) and noise (denoted by $.noise$) of the resulting ciphertext. 
    }
    \label{tbl:instructions}
    \setlength{\tabcolsep}{3pt}
      \begin{tabular}{@{}llll@{}} 
      \toprule
      Instruction & Computation & Description & Multiplicative depth   \\ 
      \midrule
      Add($ct_x$, $ct_y$) $\to ct_z$ & 
        $ct_x.data + ct_y.data$  &
        Adds two ciphertexts &
        $max(ct_x.noise, ct_y.noise)$ \\
      Add($ct$, $pt$) $\to ct_z$ & 
        $ct.data + pt.data$ &
        Adds a ciphertext and plaintext &
        $ct.noise$ \\
      Subtract($ct_x$, $ct_y$) $\to ct_z$ & 
        $ct_x.data - ct_y.data$  &
        Subtract two ciphertexts & 
        $max(ct_x.noise+ct_y.noise)$ \\
      Subtract($ct$, $pt$) $\to ct_z$ & 
        $ct.data - pt.data$ &
        Subtract a plaintext from a ciphertext &
        $ct.noise$ \\
      Multiply($ct_x$, $ct_y$) $\to ct_z$ & 
        $ct_x.data \times ct_y.data$ &
        Multiple two ciphertexts &
        $max(ct_x.noise,ct_y.noise)+1$ \\
      Multiply($ct$, $pt$) $\to ct_z$ &
        $ct.data \times pt.data$  &
        Multiply a ciphertext and plaintext &
        $ct_x.noise + 1$ \\
      Rotate($ct$, $x$)  $\to ct_z$ &  
        \begin{tabular}{@{}l@{}}$ct.data[i] \gets$ \\ \hspace{3mm} $ct.data[(i+x) mod N]$\end{tabular}  &
        Rotate a ciphertext $x$ slots to the left &
        $ct.noise$\\
      \bottomrule
    \end{tabular}
  \end{table*}
}

%% file: text/05-engine.tex
\section{Synthesis Engine}
\label{sec:synthesis_engine}

This section describes how \armadillo's synthesis engine (see \autoref{alg:synthesis_engine}) 
searches the program space (described by our \sketchname) to find an optimized HE solution
that satisfies the kernel specification.
\armadillo's synthesis engine operates by first synthesizing an initial solution. 
It then optimizes the solution by 
iteratively searching for better solutions 
until either the best program in the sketch is found 
or a user-specified time out is reached.

\armadillo's synthesis engine is a counter-example guided inductive synthesis (CEGIS) loop~\cite{sketch,oracle-synthesis}.
It uses Rosette’s built-in support for translating synthesis and 
verification queries to constraints that are solved by an SMT solver.

\subsection{Synthesizing an Initial Solution}
The first step in \armadillo's synthesis procedure is to synthesize an initial program that satisfies the user's specification.
In particular, \armadillo first attempts to complete a sketch $sketch_L$ that encodes programs using $L$ components.
Specifically, \armadillo searches for a solution $sol_0$ 
contained in $sketch_L$ that minimizes $L$ and satisfies the specification for all inputs.

We follow a synthesis procedure similar to those proposed in \cite{oracle-synthesis, gulwani2011synthesis}, and avoid directly solving the above query because it contains a universal quantifier over inputs.
Instead, we synthesize a solution that is correct for one random input rather then verify it is correct for all inputs; we then apply feedback to the synthesis query if verification fails.

\paragraph{Synthesize.} The engine starts by generating a concrete input-output example,
$(x_0, y_0)$, by evaluating the specification using a randomly generated input, $x_0$ (line 6).
The engine attempts to synthesize a program that transforms $x_0$ into $y_0$ 
by completing the sketch and finding a binding for the $L$ arithmetic instructions 
and operand holes (line 10).
We generate a synthesis query expressing $solve(sketch_L(x_0) = y_0)$, which is then compiled to constraints and solved by an SMT solver.

\paragraph{Verify.} If successful, the synthesis query described above returns a program that satisfies the input specification for the input $x_0$, but not necessarily for all possible inputs.
To guarantee that the solution is correct, \armadillo verifies the solution matches the specification for all inputs.
\armadillo leverages Rosette's symbolic evaluation and verification capabilities to solve this query. 
First, a formal specification is lifted from reference specification with symbolic execution, 
capturing the kernel's output for a bounded set of inputs as a symbolic input-output pair $(\hat{x}, \hat{y})$.
Rosette then solves the verification query $verfiy(sol(\hat{x}) = spec(\hat{x}))$.

\paragraph{Retry with Counter-example.} If verification fails, it returns a counter-example, $(x_1, y_1)$, that causes the synthesized kernel to disagree with the specification.
\armadillo then makes another attempt to synthesize a program; 
this time trying to satisfy both the initial example and counter-example. 
This process repeats until \armadillo finds a correct solution.

If the engine cannot find a solution, 
indicated when the solver returns \texttt{unsat} for any synthesis query, 
the engine concludes that 
for the given sketch, a program that implements the specification with $L$ components does not exist.
The engine tries again with a larger sketch $sketch_{L+1}$ 
that contains one more component
and this process repeats until a solution is found.
By exploring smaller sketches first, our algorithm ensures that the solution using the smallest number of components is found first.

\subsection{Optimization}
Once an initial solution is found,
\armadillo's synthesis engine attempts to improve performance
by searching for better programs contained in the sketch.
Programs are ranked according to a cost function
that \armadillo attempts to minimize.

\paragraph{Cost Function.}
\armadillo uses a cost function that multiplies the estimated latency 
and multiplicative depth of the program:
$cost(p) = latency(p) \times (1 + mdepth(p))$.
We include multiplicative depth to penalize high-noise programs,
which can lead to larger HE parameters and lower performance.

\paragraph{Cost Minimization.}
Once a solution $sol_0$ with cost $cost_0$ is found, 
we iteratively search for a new program with lower cost (line 19).
\armadillo does this by re-issuing the synthesize query with an additional constraint
that ensures a new solution $sol_1$, has lower cost: $cost_1 < cost_0$ (line 25).
This process repeats until the solver proves there is no lower cost solution and it has found the best solution or the compile time exceeds the user-specified time out.
The initial solution is only used to provide an upper-bound on cost and
is not used during the optimization synthesis queries.
This forces the engine to consider completely new programs 
with different instruction mixes and orderings.
In practice, we find that initial solutions perform well given the savings
in compile time (see \autoref{sec:synthesis_time} for discussion).

\input{figures/engine-code.tex}

\subsection{Code Generation.}

The synthesis engine outputs a HE kernel described in Quill and 
\armadillo then translates the Quill program into a SEAL program~\cite{sealcrypto}.
SEAL is a HE library that implements the BFV scheme.
Quill instructions map directly to SEAL instructions, 
so this translation is simple, but the code generation handles a few post-processing steps.
For example, \armadillo inserts special \textit{relinearization} instructions after each ciphertext-ciphertext multiplication.
Relinearization does not affect the results of the HE program but is necessary to handle ciphertext multiply complexities.

%% file: figures/engine-code.tex
\newcommand\linecomment[1]{\textcolor{gray}{\textit{#1}}}
\algrenewcommand\algorithmicindent{1.0em}
\algblock{Input}{EndInput}
\algnotext{EndInput}
\algblock{Output}{EndOutput}
\algnotext{EndOutput}
\newcommand{\Desc}[2]{\State \makebox[5em][l]{#1}#2}

\begin{algorithm}[t]
   \caption{Synthesis engine}
   \label{alg:synthesis_engine}
   \algtext*{EndWhile}% Remove "end while" text
   \algtext*{EndIf}% Remove "end if" text
   \algtext*{EndFunction}% Remove "end function" text

   \begin{algorithmic}[1]

      \Input
         \Desc{$spec$}{Kernel reference program}
         \Desc{$sketch$}{Partial HE program}
      \EndInput

      \State \linecomment{Synthesize first solution}
      \Function{synthesize}{}
         \State $y_0 \gets spec(x_0)$ \Comment{Random input-output example}
         \State $\hat{y} = spec(\hat{x})$ \Comment{Symbolic input-output}
         \State examples = [$(x_0, y_0)$]
         \While {true}
            \State sol $\gets$ solve(sketch s.t. $y$=sketch({$x$}))
            \If {sol is \texttt{unsat}}
               \State \textbf{return} \texttt{False} \Comment{Sketch too restrictive}
            \EndIf
            \State cex $\gets$ verify($\hat{y} = solution(\hat{x})$)
            \If {cex = \texttt{unsat}}
               \State \textbf{return} sol
            \EndIf
            \State $(x, y)$ $\gets$ extract(cex) \Comment{Get counterexample}
            \State examples.append($(x, y)$)
         \EndWhile
      \EndFunction

      \State \linecomment{Minimize cost}
      \Function{optimize}{}
         \State $sol$ $\gets$ synthesize()
         \State $c'$ $\gets$ cost(sketch)
         \State $sol'$ $\gets$ $sol$
         \While {$sol'$ is \texttt{sat}}
            \State $c$ $\gets$ cost($sol$), $sol$ $\gets$ $sol'$
            \State $sol'$ $\gets$ solve(sketch  s.t. $y$=sketch({$x$}) \& $c' < c$)
            \State \linecomment{$<$verify $sol'$ and add cex if needed$>$}
         \EndWhile
         \State \textbf{return} $sol$
      \EndFunction

   \end{algorithmic}
\end{algorithm}

%% file: text/06-optimizations.tex
\section{Synthesis Formulation Optimizations}
\label{sec:scaling_up}

Scaling \armadillo to handle larger kernels requires optimizing the synthesis formulation.
Since the search space grows super exponentially, 
it quickly becomes intractable---a five instruction HE program can have millions of candidate programs.
This section describes optimizations developed 
to scale up our formulation and their impact on the results.

\subsection{Rotation Restrictions}

HE Rotation instructions are used to align different vector slots 
within a ciphertext to perform computation such as reductions.
Ciphertext slots can be rotated by up to $n$, 
the size of the ciphertext vector, 
which introduces a large number of possible rotations for the synthesizer to select from.
In practice, we observe that of all possible rotations only a few patterns are ever used.
For example, in our $G_x$ kernel each output element only depends on its neighbors in the 3$\times$3 window, implying rotations that align input elements from outside this window are not necessary.
By restricting rotations, we can scale up the synthesis process by pruning away irrelevant potential programs.

To optimize for this, 
we introduce two types of rotation restrictions for tree reductions and sliding windows.
For sliding window kernels, which are commonly used in image processing, 
we use the restriction described above to restrict rotation holes \texttt{??rot} 
to align elements covered by the window.
The tree reduction restricts rotations to powers of two and is used for kernels that implement an internal reduction within the ciphertext.
For example, in a dot product elements in the vector are summed to produce one value.
Restricting the rotations to powers of two constrains the output programs 
to perform the summation as a reduction tree.

\subsection{Constraint Optimizations}

We also apply a number of common constraint optimization techniques to improve synthesis speed and scalability.
We employ symmetry breaking to reduce the search space for add, multiply, and rotate.
For example, the programs $a+b$ and $b+a$ are functionally equivalent
but appear as two unique solutions to a solver.
Restricting operands to occur in increasing order 
eliminates redundant candidate solutions and improves synthesis speed.
For rotations we impose symmetry breaking by forcing only left rotations,
since a left rotation by $x$ is equivalent to a right rotation by $n-x$.
We also enforce solutions use static single assignment 
to instill an ordering and break symmetries between programs that are functionally equivalent but write to different destination ciphertexts.

Our synthesis formulation also uses restricted bitwidth instead of full precision bit vectors
to reduce the number of underlying variables the solver needs to reason about.
Ordinarily, the number of solver variables scales linearly with bitwidth; 
however, we do not need the bit accurate behavior,
only the operator functionality, so this optimization does not affect correctness of the solution.

\subsection{Multi-step synthesis}

One of the limitations of program synthesis is its inability to scale to large kernels~\cite{gulwani-program-synthesis}.
With the above optimizations, \armadillo scales to roughly 10-12 instructions, 
but beyond that the program space becomes intractable to search.
Many applications in image processing, neural networks, and machine learning have natural break points. 
For instance, an image processing pipeline may have cascaded stencil computations for sharpening, blurring, and edge detection which have natural boundaries.
To scale beyond the limitations of program synthesis, 
we leverage these natural breakpoints to 
partition larger programs into segments and synthesize them independently.
In \autoref{sec:evaluation}, we show how this partitioning into a multistep synthesis problem can allow \armadillo to scale to longer kernels. 

%% file: text/07-evaluation.tex
\section{Evaluation}
\label{sec:evaluation}

This section evaluates \armadillo's synthesized programs and compares them against expert-optimized baselines (see \autoref{sec:synthesis_results}).
We also report how long \armadillo takes to synthesize kernels (see \autoref{sec:synthesis_time}).
We find that \armadillo is able to synthesize a variety of kernels that are at least as good or better than an expert-written version, and in most cases can synthesize a kernel in under a few minutes.

\begin{figure}[t]
    \includegraphics[width=0.48\textwidth]{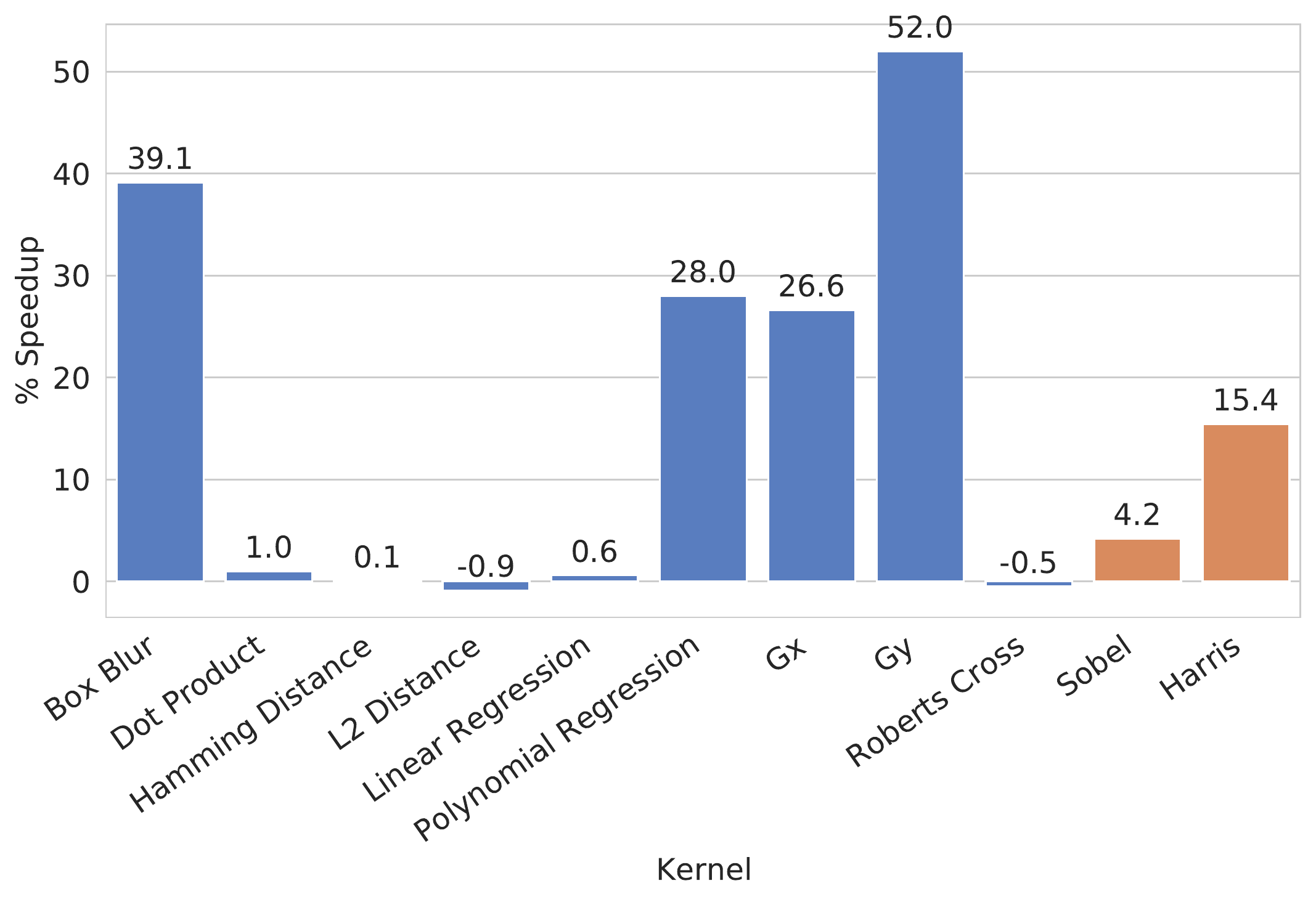}
    \caption{Speedup of \armadillo synthesized kernels compared to the baseline, results are averaged over 50 runs. 
    Kernels in blue are directly synthesized while kernels in orange use multi-step synthesis.}
    \label{plot:latency_plot}
\end{figure}

\subsection{Methodology}

\begin{table}[t]
    \caption{A comparison of instruction count and computation depth 
    of baseline and synthesized kernels.}
    \begin{tabular}{lcc|cc}
        \toprule
        Kernel & \multicolumn{2}{c}{Baseline} & \multicolumn{2}{c}{Synthesized} \\
         & Instr. & Depth & Instr. & Depth \\
        \midrule
        Box Blur & 6 & 3& \textbf{4} & 4\\
        Dot Product & 7 & 7 & 7 & 7\\
        Hamming Distance & 6 & 6 & 6 & 6\\
        L2 Distance & 9 & 9 & 9 & 9\\
        Linear Regression & 4 & 4 & 4& 4\\
        Polynomial Regression & 9 & 6 & \textbf{7} & 5\\
        Gx & 12 & 4 & \textbf{7} & 6\\
        Gy & 12 & 4 & \textbf{7} & 6\\
        Roberts Cross & 10 & 5 & 10 & 5\\
        Sobel & 31 & 7 & \textbf{21} & 9\\
        Harris & 59 & 14 & \textbf{43} & 17\\
        \bottomrule
    \end{tabular}
    \vspace{3ex}
    \label{tbl:instruction_count}
\end{table}

\armadillo is implemented with Rosette v3.1~\cite{rosette},
and configured to use Boolector~\cite{boolector} 
as its backend SMT solver.
Synthesized kernels are compiled down to SEAL v3.5's BFV library~\cite{sealcrypto}.
We time out all synthesis jobs after 20 minutes of no progress and return the current best solution.
When running \armadillo's kernels, security parameters are set to guarantee a 128-bit security level; both baseline and synthesized kernels use the same settings. 
All experiments are conducted on a 3.7 GHz Intel Xeon W-2135 CPU with 64 GB of memory.

\paragraph{Workloads.} 
We evaluate \armadillo using common kernels found in
linear algebra, machine learning, and image processing listed in \autoref{tbl:syntime}.
Since there is no standardized benchmark for compiling HE kernels,
we attempt to be as diverse and representative
in our selection as possible.
For example, dot product, L2 distance, and linear and polynomial regression kernels are building blocks of machine learning applications, while the x/y-gradient ($G_x$/$G_y$) and Roberts cross kernels are used in image processing applications.

Kernels are modified to omit operations not directly supported by HE.
For instance, the canonical L2 distance kernel uses a square root,
but many applications (e.g., k-nearest neighbors) can use squared distance with negligible effect on accuracy~\cite{flann}.
Finally, because BFV cannot implement data-dependent branches or conditionals, 
applications that require these operations are calculated up to a branch.
For example, our Harris corner detector implementation returns an image of response values that the user must decrypt and apply a threshold over to detect the corners.

\paragraph{Baselines.} 
We compare \armadillo's code quality against
an expert's hand-written implementation that seeks to minimize logic depth.
Logic depth minimization was chosen to reflect the state-of-the-art solution
that was recently proposed for optimizing HE kernels
under boolean HE schemes~\cite{lee2020optimizing}.
The paper suggests that optimizing logical computation depth also minimizes noise,
as fewer successive operations intuitively compounds less noise between any input-output.
To minimize depth, these programs attempt to perform as much computation as possible in early levels
of the program and implement all reductions as balanced trees.
In addition, all our baseline implementations use packed inputs (i.e., are not scalar implementations) to minimize latency.

\subsection{Synthesized Kernel Quality}
\label{sec:synthesis_results}
To understand the quality of \armadillo's synthesized programs, we compare
instruction count, program depth, and run time against the
hand-optimized baseline.
We report run time speedups in \autoref{plot:latency_plot}, with all times averaged over 50 independent runs and instruction counts in \autoref{tbl:instruction_count}.

The results show that \armadillo's kernels have
\textit{similar or better performance}
compared to the hand-written baselines.
For some kernels such as dot product, L2 distance, and Roberts cross,
\armadillo generates roughly the same kernel as the hand-written implementation.
The synthesized and baseline implementations may have different orderings of independent instructions, resulting in small performance differences.
We examined the kernels and found this is because the optimal programs for
depth (baseline) and instruction count (\armadillo) are the same.

For more complex kernels ($G_x$, $G_y$), polynomial regression, and box blur),
we observe \armadillo's programs have notably better run times, up to \speedupbest and use fewer instructions.
Our speedups are a result of \armadillo being able to identify different types of optimizations.
For example, our synthesized polynomial regression kernel found an algebraic optimization that factored out a multiplication
similar to $ax^2+bx = (ax + b)x$, resulting in a kernel that used 7 instructions instead of 9 and was \speeduppoly faster than the baseline.
We analyze more of these optimizations in \autoref{sec:tradeoffs}.

For these kernels,
each handwritten baseline took on the order of a few hours to a day to implement, debug, and verify;
for a non-expert unfamiliar with HE and SEAL, this would take much longer.
The results show that \armadillo can effectively automate the
tedious, time-consuming task of handwriting these kernels without sacrificing quality.

\paragraph{Multi-step Synthesis Evaluation. }
We also used \armadillo's synthesized kernels to compile larger HE applications.
Specifically, \armadillo's $G_x$ and $G_y$ kernels are used to implement the Sobel operator, and $G_x$, $G_y$, and box blur kernels were used to implement the Harris corner detector, shown in orange in \autoref{plot:latency_plot}. 
By leveraging \armadillo synthesized kernels, our Sobel operator and Harris corner detector were \speedupsobel and \speedupharris faster than the baseline, and used 10 and 16 fewer instructions respectively.
These results show that we can speedup larger applications
by synthesizing the core computational kernels these applications rely on.

\subsection{Analysis of Synthesized Kernels}
\label{sec:tradeoffs}

\begin{figure}[t]
    \centering
    \includegraphics[width=0.95\linewidth]{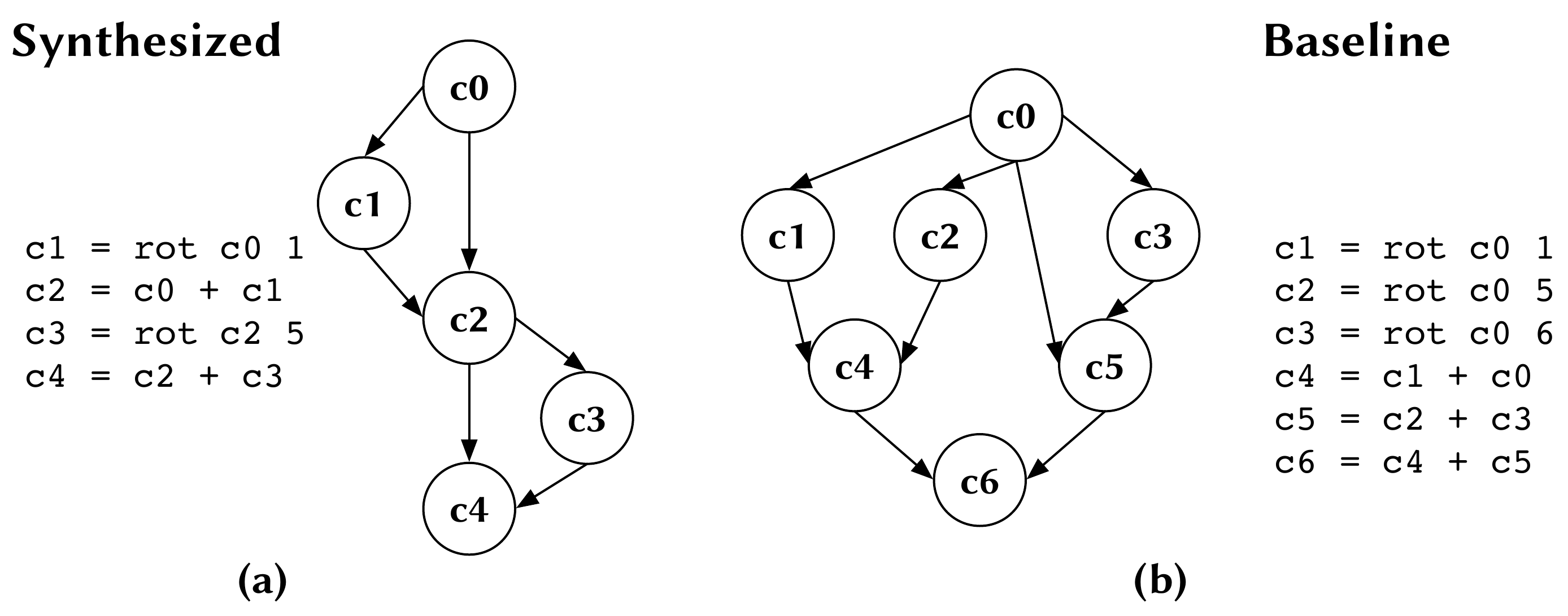}
    \caption{HE kernels for box blur. (a) Synthesized kernel with minimal number of instruction (b) Hand-optimized minimal depth kernel.
    Porcupine achieves a much higher performing kernel by separating kernels
    and use fewer instructions
    which, even though the logical depth increases, results in a 39\% speedup.}
    \label{fig:boxblur_dataflow}
    \vspace{2ex}
\end{figure}

We now analyze the synthesized and baseline implementations of the box blur and $G_x$ kernels to demonstrate the trade-offs explored by \armadillo.
\autoref{fig:boxblur_dataflow} compares \armadillo's and the baseline's box blur.
The baseline implements this kernel in six instructions with three levels of computation.
In the first level, elements are aligned in the window with rotations and then summed in a reduction tree.
\armadillo's synthesized kernel uses four instructions with five levels; decomposing the 2D convolution into two 1D convolutions to perform the same computation with fewer instructions.
Furthermore, despite having a greater logical depth, the synthesized solution consumes the same amount of noise as the baseline.
By focusing on minimizing depth, the baseline misses the separable kernel optimization because it was not the minimum depth solution.

\begin{figure}[t]
\centering
\includegraphics[width=0.95\linewidth]{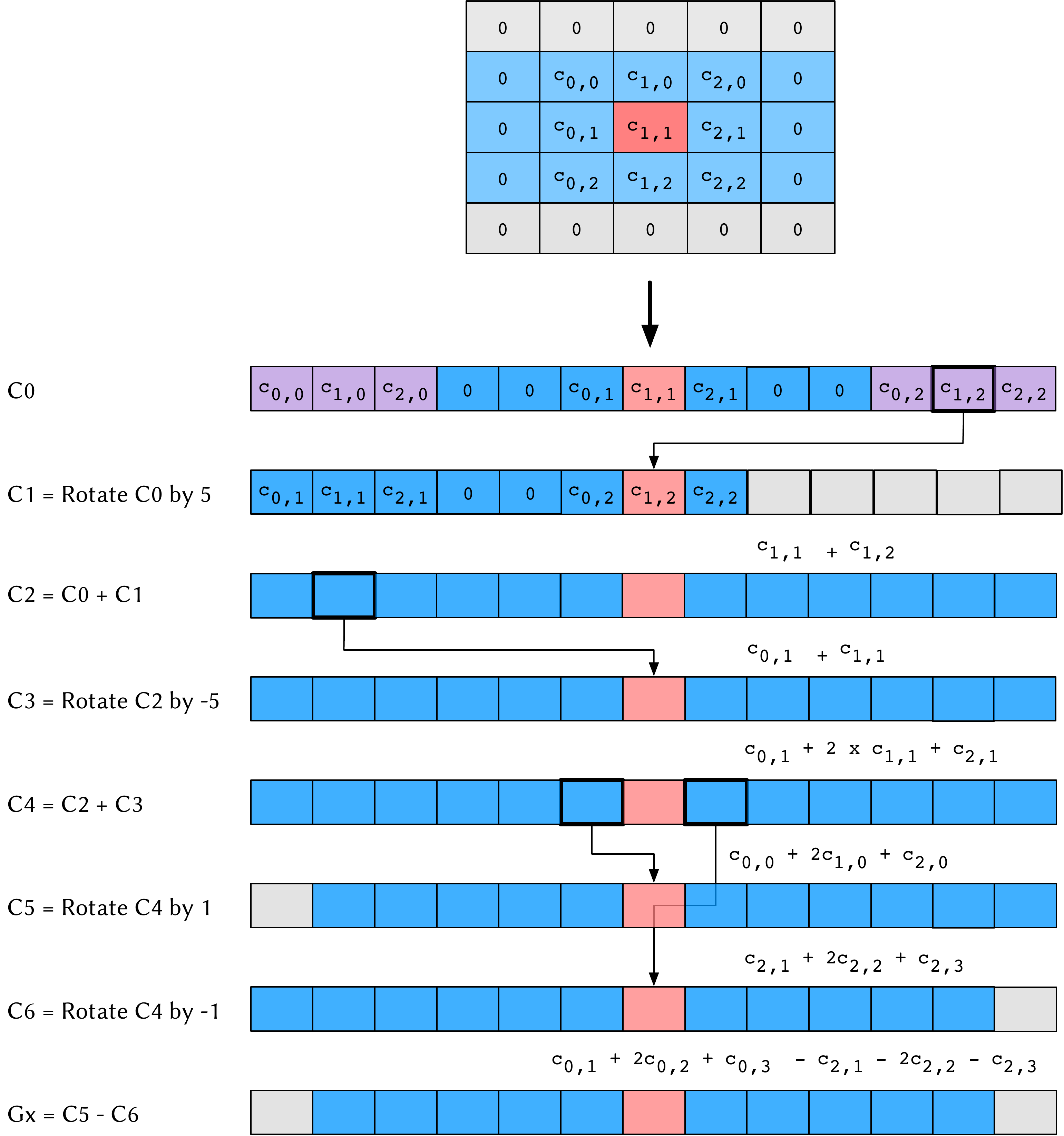}
\vspace{2ex}
\caption{\armadillo optimized $G_x$ kernel. An optimized implementation packs the entire image into one ciphertext and schedules computation with rotations. Purple slots contain elements that are used to compute the final red slot. The value contained in the red slot are tracked on the right hand side.}
  \label{fig:gx_layout}
  \vspace{2ex}
\end{figure}

We observe similar results for the $G_x$ kernel and show the synthesized and baseline programs in \autoref{fig:gx_dataflow}.
The depth-optimized baseline applies the same strategy as the box blur kernel, first aligning elements in the sliding window then combining them in a balanced reduction tree.
The $G_x$ kernel weights some of the neighbor elements by two, and the baseline substitutes the multiplication with a cheaper addition (operand c11 in \autoref{fig:gx_dataflow}b).
The synthesized $G_x$ kernel has a very different program structure from the baseline.
\armadillo discovers the filter is separable and decomposes the kernel into two 1D filters, requiring a different set of rotations and schedule to implement correctly as depicted in \autoref{fig:gx_layout}.
\armadillo's synthesized solutions automatically also substitutes the multiplication by 2 with an addition which is performed at c4 in parallel with other additions.

{
\begin{figure}[t]
\centering
\includegraphics[width=0.98\linewidth]{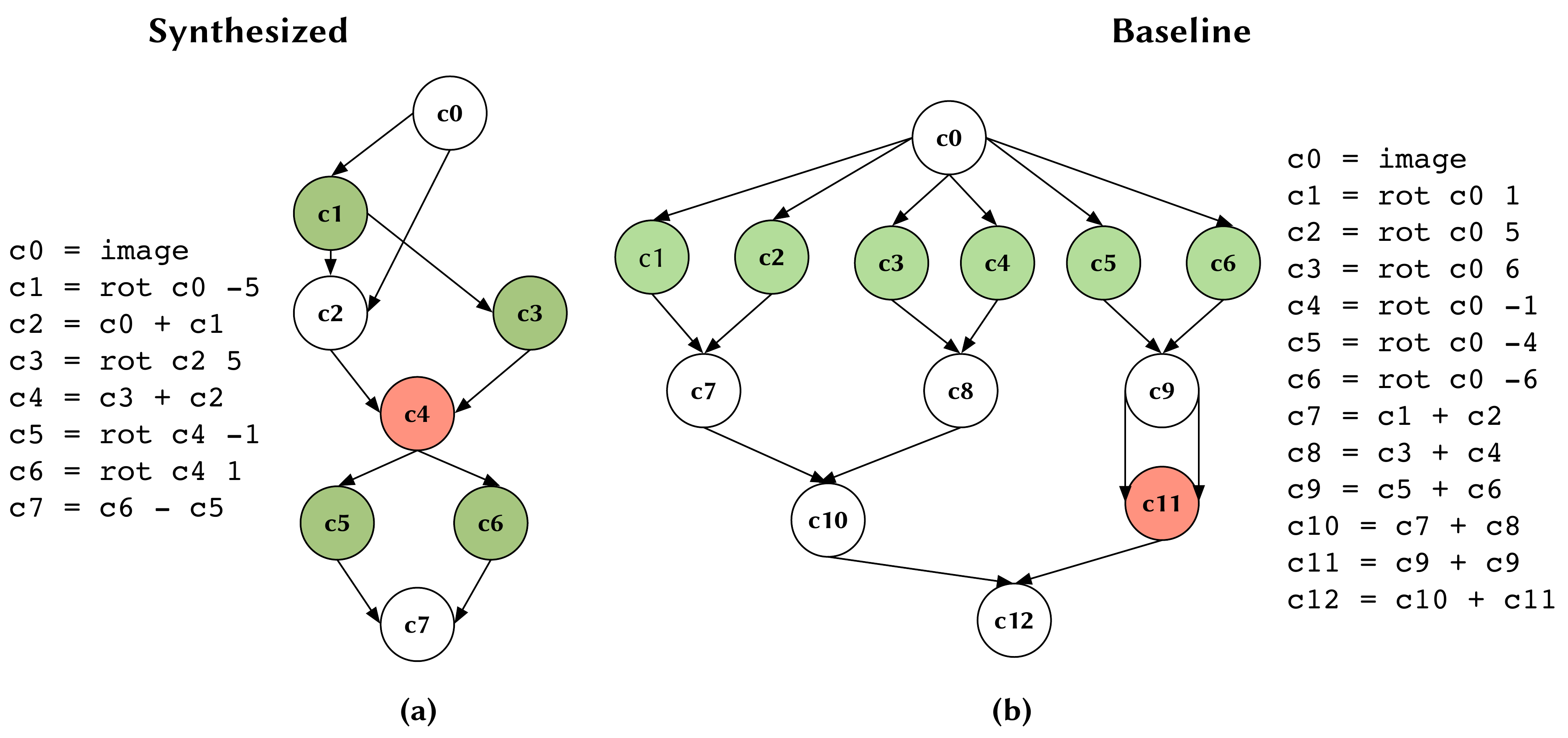}
\caption{(a) Synthesized and (b) baseline $G_x$ kernel. The synthesized kernel uses 7 instructions while the baseline uses 12 instructions. The synthesized kernel optimizes the computation to separate the 2D convolution into two 1D convolutions and interleaves rotation and computation. Ciphertexts generated by rotations are marked in green and the ciphertext where multiplication by 2 is implemented with an addition is in red.}
\label{fig:gx_dataflow}
\end{figure}
}

While minimizing for logical depth is a good guideline for minimizing noise in scalar HE programs, our results show it is not guaranteed to find the optimal implementations for vector HE constructions,
like BFV, and can leave significant unrealized performance (e.g., up to \speedupbest for box blur).
Because \armadillo searches for vectorized implementations, and tracks program latency and multiplicative depth
it can outperform the heuristic for programs with more complex dependencies.

\begin{table*}[th]
    \caption{Synthesis time and number of examples used by \armadillo. Initial time is the time to synthesize a solution and total time includes time spent optimizing. Reported values come from the median of 3 runs.}
    \label{tbl:syntime}
    \input{data/syntable.tex}
    \vspace{2ex}
\end{table*}

\subsection{Synthesis Time}

\label{sec:synthesis_time}

\autoref{tbl:syntime} reports the time it took to synthesize each kernel averaged over three runs.
We report how long it took to find an initial solution
and the cumulative time it took to find an optimized solution.
For most of the kernels we were able to synthesize an initial solution in under 30 seconds and synthesize an optimized solution under 2 minutes. 
The Roberts cross kernel required more time, taking over 2 minutes to synthesize an initial solution and in total to 27 minutes to optimize.
This is because the Roberts cross kernel required a sketch with 8 instructions, which took longer to search over.
Additionally, the optimization phase of the synthesis engine must prove it found the best solution contained in the sketch, requiring the SMT solver explore the entire search space.

In terms of input-output examples required by \armadillo during the synthesis process, we typically only require one example to synthesize a solution;
however, for some kernels such as Hamming distance we required up to 6 input-output examples be generated during synthesis.
We find kernels that produce a single-valued output, like Hamming distance, require
more examples than kernels that produce a vector output (e.g., image processing kernels).
This is because the synthesis engine can find many input-dependent (but not general) programs.

\paragraph{Cost Trajectory.}
\autoref{tbl:syntime} also reports the cost of the initial and final solutions found by \armadillo.
For some kernels, the initial and first solution
\armadillo finds are the same.
This indicates that there was only one correct solution in the minimum $L$-sized sketch, or that \armadillo found the best solution on the first try.

The time between \armadillo reporting the initial and final solution is spent proving that it found the best solution in the sketch.
After the initial solution is found, users can terminate \armadillo early to shorten compile times.
While this does not guarantee the best solution was found, it will minimize arithmetic instructions.

\paragraph{Analysis of local rotate sketches.}
In \autoref{sec:sketch} we introduced our local rotate sketches treat rotations as instruction operands instead of components.
We could have alternatively required users explicitly add rotations to the list of components supplied in the sketch (which we refer to as \textit{explicit rotation} sketches). 
However, explicit rotation sketches describe a larger space of programs that includes the space described by our \sketchname sketches.

In small kernels, e.g., box blur, the synthesis time using \sketchname sketches was faster than the explicit rotation sketch; the explicit rotation sketch
took only 3 seconds to synthesize verses 10 seconds when using a
\sketchname sketch.
However, when we search over larger programs the explicit rotation sketch scales poorly.
Synthesizing the $G_x$ kernel took over 400 seconds to find an initial solution then over 30 minutes total when using the explicit rotation sketch.
On the other hand the \sketchname sketches found the same solution in about 70 seconds seconds, showing that \sketchname does improve synthesis scalability and search time.

%% file: data/syntable.tex
\begin{tabular}{lrrrrr}
\toprule
        Kernel & Examples & Initial Time (s) & Total Time (s) & Initial Cost & Final Cost\\
    \midrule
              Box Blur &        1 &             1.99 &           9.88 &      1182 &        592 \\
           Dot Product &        2 &             1.27 &          15.16 &      1466 &       1466 \\
      Hamming Distance &        3 &             0.87 &           2.24 &      1270 &        680 \\
           L2 Distance &        2 &            27.57 &         114.28 &      1436 &       1436 \\
     Linear Regression &        2 &             0.50 &           0.69 &       878 &        878 \\
 Polynomial Regression &        2 &            24.59 &          47.88 &      2631 &       2631 \\
                    Gx &        1 &            14.87 &          70.08 &      1357 &        975 \\
                    Gy &        1 &             9.74 &          49.52 &      1773 &        767 \\
          Roberts Cross &        1 &           212.52 &         609.64 &      2692 &       2692 \\
\bottomrule

\end{tabular}

%% file: text/08-related-work.tex
\section{Related Work}
\label{sec:related_work}

\subsection{Compilers for Homomorphic Encryption}

Recent work proposes domain-specific and general compilers for HE~\cite{chet, ngraph, eva, ckks, ramparts, cingulata}.
Prior work such as CHET~\cite{chet} and nGraph-HE~\cite{ngraph} are domain-specific HE compilers for deep neural networks (DNNs).
CHET optimizes the data layout of operations in DNNs while nGraph-HE added an HE extension to an existing DNN compiler with new graph-level optimizations.
Cingulata~\cite{cingulata} and Lobster~\cite{lee2020optimizing} only target Boolean HE constructions and propose compilation strategies that rely on multiplicative depth minimization and synthesizing rewrite rules.
EVA~\cite{eva} automatically inserts low-level HE instructions such as rescale and modswitch using custom rewrite rules but requires a hand-crafted HE kernel as input.
\armadillo supports vectorized HE constructions, generalizes to non-DNN computation, and is more general than the rewrite rule-based prior work since program synthesis automates some of this reasoning.

The closest work to ours is Ramparts~\cite{ramparts} which is a HE compiler that translates plaintext Julia programs to equivalent HE implementations. Unlike \armadillo, Ramparts does not support packed vectorization (i.e., one task cannot use multiple slots in a ciphertext) which is required for taking advantage of SIMD parallelism within a task and improving latency. 
In contrast, \armadillo supports packed data inputs and can generate kernels with rotations.
Furthermore, Ramparts relies on the structure of the input Julia program to serve as the seed for symbolic execution-based methodology which produces a computational circuit that is optimized and lowered to HE instruction with rewrite rules. In contrast, \armadillo places essentially no constraints on the structure of the programs it synthesizes other than the number of instructions it can contain. This enables \armadillo to consider a wider range of programs when optimizing. 

Overall, \armadillo is the only compiler that supports the task of automatically scheduling and mapping packed vectorized HE kernels for general computation.

\subsection{Compilers for Privacy-Preserving Computation}

Compiler support has also been proposed for other privacy-preserving techniques, such as differential privacy (DP)~\cite{dwork2006calibrating} and secure multi-party computation (MPC)~\cite{goldreich2019play, yao1986generate} to automatically enforce or reason about restrictions and constraints by these technologies.
For instance, DP requires adding noise to the algorithm and deriving that the effect of an individual's information is in fact differentially private (i.e., has indistinguishable effect on the aggregate data).
In DP, there are proposals for using type systems to enforce differential privacy~\cite{gaboardi2013linear, near2019duet, reed2010distance}.
Other programming language techniques~\cite{barthe2016programming} include dynamic approaches~\cite{mcsherry2009privacy, mcsherry2010differentially, roy2010airavat}, static checking~\cite{reed2010distance, palamidessi2012differential, gaboardi2013linear}, and machine-checked proofs~\cite{barthe2012probabilistic}.
A similar trend is occurring in MPC where implementations must also comply with design constraints to collaboratively compute functions while still protecting private inputs from other users.
Recent work by ~\cite{zahur2015obliv, wang2016emp, songhori2015tinygarble, rastogi2014wysteria, hastings2019sok} proposes and/or evaluates general-purpose compiler for MPC.

\subsection{Synthesizing Kernels}
Prior work has also used program synthesis to generated optimized kernels for other targets.
For example, Spiral~\cite{puschel2005spiral} generates optimized DSP kernels using both inductive and deductive synthesis techniques.
Swizzle Inventor ~\cite{swizzle-inventor} synthesized optimized data movement for GPU kernels from a sketch that specified that computation strategy and left data movement unspecified. 
Because their objective was to only optimize data movement, they relied on canonicalization for verification (not an SMT solver) which does not allow their synthesis formulation to optimize algebraic expressions but improves synthesis time.
On the other hand, our synthesis formulation needs to optimize algebraic expressions as part of selecting arithmetic instructions so requires an SMT solver.

%% file: text/09-conclusion.tex
\section{Conclusion}
\label{sec:conclusion}

We presented \armadillo, a program synthesis-based compiler that automatically generates vectorized HE kernels.
\armadillo automatically performs the instruction selection and scheduling to generate efficient HE kernels and minimize the HE noise budget.
By automating these tasks, \armadillo abstracts away the details of constructing correct HE computation so that application designers can concentrate on other design considerations.
HE is still a rapidly maturing area of research and there is limited related work in this space.
As a result, we expect that in future work we will see rapid improvements to compilation infrastructure such as ours.